\documentstyle[twoside, 12pt]{article}

\newtheorem{remark}{Remark}

\newtheorem{assumption}{Assumption}
\def\qed{ \ \vrule width.2cm height.2cm depth0cm\smallskip}

\newcommand{\brm}{\begin{rem}}
\newcommand{\ermq}{\end{rem}}

\newcommand{\ba}{\begin{array}}
\newcommand{\ea}{\end{array}}
\newcommand{\be}{\begin{equation}}
\newcommand{\ee}{\end{equation}}
\newcommand{\bea}{\begin{eqnarray}}
\newcommand{\eea}{\end{eqnarray}}
\newcommand{\beaa}{\begin{eqnarray*}}
\newcommand{\eeaa}{\end{eqnarray*}}

\def \R{I\!\!R}

\def\si{\sigma}
\def\t{\tau}
\def\f{\varphi}
\def\th{\theta}

\def\cF{{\cal F}}

\def\cT{{\cal T}}

\def\no{\noindent}

\def\ms{\medskip}
\def\bs{\bigskip}
\def\q{\quad}

\def\cd{\cdot}
\def\cds{\cdots}

\def\bF{{\bf F}}

\def\qed{ \hfill \vrule width.25cm height.25cm depth0cm\smallskip}
\newcommand{\dfnn}{\stackrel{\triangle}{=}}
\newcommand{\basa}{\begin{assumption}}
\newcommand{\easa}{\end{assumption}}

\newcommand{\bas}{\begin{assum}}
\newcommand{\eas}{\end{assum}}

\def\esssup{\mathop{\rm esssup}}
\def\essinf{\mathop{\rm essinf}}

 \def\cd{\cdot}
\def\cds{\cdots}

\def\dis{\displaystyle}

\def\bF{{\bf F}}

\newtheorem{thm}{Theorem}[section]
\newtheorem{lem}[thm]{Lemma}

\newtheorem{rem}[thm]{Remark}

\newtheorem{defn}[thm]{Definition}
\newtheorem{assum}[thm]{Assumption}

\title{The Continuous Time Nonzero-sum Dynkin Game Problem and Application in Game Options}
\author{Said
Hamad\`ene\thanks{Universit\'e du Maine, D\'epartement de
Math\'ematiques, Equipe Statistique et Processus, Avenue Olivier
Messiaen, 72085 Le Mans, Cedex 9, France. e-mail:
hamadene@univ-lemans.fr}\,\,\ and \,\,Jianfeng Zhang\thanks{USC
Department of Mathematics, 3620 S. Vermont Ave, KAP 108, Los Angeles, CA
90089, USA. e-mail:jianfenz@usc.edu. Research supported in part by
NSF grants DMS 04-03575 and DMS 06-31366. Part of the work was done
while this author was visiting Universit\'e du Maine, whose
hospitality is greatly appreciated.}}
\begin{document}
\date{\today}
\maketitle

\begin{abstract}{\it In this paper we study the nonzero-sum
Dynkin game in continuous time which is a two player non-cooperative
game on stopping times. We show that it has a Nash equilibrium point
for general stochastic processes. As an application, we consider the
problem of pricing American game contingent claims by the utility
maximization approach.}
\end{abstract}
{\bf AMS Classification subjects}: 91A15; 91A10; 91A30; 60G40; 91A60.
\medskip

\no {$\bf Keywords$}: Nonzero-sum Game; Dynkin game; Snell
envelope; Stopping time; Utility maximization; American game
contingent claim. \ms

\vfill\eject

\section{Introduction}
\setcounter{equation}{0}

Dynkin games of zero-sum or nonzero-sum, continuous or discrete time
types, are games on stopping times. Since their introduction by E.B.
Dynkin in \cite{dynkin}, they have attracted a lot of research
activities (see e.g. \cite{nazaret2, alvarez, bfried, bfried2,
bismut, catlep, CK, ekstrompeskir, ekstromvilleneuve, etourneau,
hamadene, ls, lepmaing, morimoto1, morimoto2, nagai, oht, stet, tv}
and the references therein). \ms

To begin with let us describe briefly those game problems. Assume we
have a system controlled by two players or agents $a_1$ and $a_2$.
The system works or is alive up to the time when one of the agents
decides to stop the control at a stopping time $\t_1$ for $a_1$ and
$\t_2$ for $a_2$. An example of that system is a {\it recallable
option} in a financial market (see \cite{hamadene, kifer} for more
details). When the system is stopped the payment for $a_1$ (resp.
$a_2$) amounts to a quantity $J_1(\t_1,\t_2)$ (resp. $J_2(\t_1,\t_2)$)
which could be negative and then it is a cost. We say that the
nonzero-sum Dynkin game associated with $J_1$ and $J_2$ has a Nash
equilibrium point (NEP for short) if there exists a pair of stopping
times $(\t_1^*,\t_2^*)$ such that for any $(\t_1, \t_2)$ we have:
$$
J_1(\t_1^*,\t_2^*)\geq J_1(\t_1,\t_2^*) \mbox{ and } J_2(\t_1^*,\t_2^*)\geq
J_2(\t_1^*,\t_2).$$ The particular case where $J_1+J_2=0$ corresponds
to the zero-sum Dynkin game. In this case, when the pair
$(\t_1^*,\t_2^*)$ exists it satisfies
$$
J_1(\t_1^*,\t_2)\leq J_1(\t_1^*,\t_2^*)\leq J_1(\t_1,\t_2^*), \mbox{ for
any }\t_1,\,\t_2.$$ We call such a $(\t_1^*,\t_2^*)$ a saddle-point for
the game. Additionally this existence implies in particular that:
$$\inf_{\t_1}\sup_{\t_2}J_1(\t_1,\t_2)=\sup_{\t_2}\inf_{\t_1}J_1(\t_1,\t_2),$$
$i.e.$, the game has a value. \ms

Mainly, in the zero-sum setting, authors aim at proving existence of
the value or/and a saddle point for the game while in the
nonzero-sum framework they focus on the issue of existence of a NEP
for the game. \ms

In continuous time, for decades there have been a lot of works on
zero-sum Dynkin games \cite{nazaret2,alvarez, bfried2,  bismut, CK,
dynkin, ekstrompeskir, ekstromvilleneuve,  hamadene, ls, lepmaing,
morimoto1, stet, tv}. Recently this type of game has attracted a new
interest since it has been applied in mathematical finance (see e.g.
\cite{bk, hamadene, kkal, kifer}) in connection with the pricing of
American game options introduced by Y.Kifer in \cite{kifer}.
Comparing with the zero-sum setting, there are much less results on
nonzero-sum Dynkin games in the literature.
%have been less considered and consequently the literature on this subject is fairly less important.
Nevertheless in the Markovian framework, among other papers, one can
quote \cite{bfried, catlep, nagai, oht} which deal with the
nonzero-sum Dynkin game. %If the framework is not markovian,
%to our best knowledge, the most general work on this field is by
In non-Markovian framework, E.Etourneau \cite{etourneau} showed that the game has a
NEP if some of the processes which define the game ($Y^1$ and $Y^2$
of (\ref{J12}) below) are supermartingales. Note that even in the
Markovian setting, an equivalent condition is supposed. On the other hand,
there are some other works which study the existence of approximate equilibrium points
(see e.g. \cite{morimoto1}).

The main objective of this work is to study the existence of NEP
for nonzero-sum Dynkin games in non-Markovian framework. For very general processes,
 we construct an NEP and thus it always exists. This removes the Etourneau's type of
conditions and, to our best knowledge, is novel in the literature.
Our approach is based on the Snell envelope theory.
We next apply our general existence result to price American Game
Contingent Claim by the utility maximization approach. Kuhn \cite{kuhn2}
studied a similar problem by assuming that the agents $a_1$ and $a_2$ use
only discrete stopping times and exponential utilities. We remove these constraints.
%Our result
%completes another one on the same subject by C.Kuhn \cite{kuhn2}
%where he assumed that the agents $a_1$ and $a_2$ are allowed to use
%only discrete stopping times and exponential utilities. In this article,  $\qed$
%, and this is its novelty, is to
%remove this latter condition and to show that the nonzero-sum Dynkin
%game has a NEP for general processes. \ms

The rest of the paper is organized as follows. In Section 2, we precise the
setting of the problem and give some preliminary results related to
the Snell envelope notion. In Section 3, we construct a sequence of
pairs of decreasing stopping times and show that their limit pair
%which converge to a limit which
%we show further that it
is  an NEP for the game. Finally in
Section 4, we apply the result of Section 3 to price American Game
Contingent Claim by the utility maximization approach.
%Our result
%completes another one on the same subject by C.Kuhn \cite{kuhn2}
%where he assumed that the agents $a_1$ and $a_2$ are allowed to use
%only discrete stopping times and exponential utilities. In this
%article, we remove these conditions.
$\qed$

\section{Formulation of the problem}
\setcounter{equation}{0}
Throughout this paper $T$ is a real
positive constant which stands for the horizon of the problem and
$(\Omega ,\cF, P)$ is a
 fixed probability space on which is defined a filtration ${\bF}:=(\cF_{t})_{0\le t\leq T}$
 which satisfies the usual conditions, $i.e.$, it is complete and right continuous.
\ms

Next:

%
%- let $p>1$ be a real constant and $\cS^p$ be the set of
%$\cF$-adapted continuous and $\R$-valued processes $(\zeta_t)_{t\leq
%T}$ such that $E[\sup_{t\leq T}|\zeta_t|^p]<\infty$
%
- for any $\bF$-stopping times $\th$, let $\cT_\th$ denote the set of
 $\bF$-stopping times $\tau$ such that $\tau \in [\th,T]$, P-a.s.
 % and,
%on the other hand, $E_t[.]$ the conditional expectation $w.r.t.$
%$\cF_t$, i.e., $E_t[A]:=E[A|{\cal F}_t]$, $\forall A\in \cF$.

- let [D] denote the space of $\bF$-adapted $\R$-valued right
continuous with left limits (RCLL for short) processes $\zeta$ such
that the set of random variables $\{\zeta_\tau, \,\,\tau \in{\cal
T}_0\}$ are uniformly integrable.
\medskip

We consider a game problem with two players $a_1$ and $a_2$. For $i=1, 2$, the player $a_i$ can choose a stopping time $\t_i \in \cT_0$ to stop the game. So the game actually ends at $\t_1\wedge \t_2$. Each player $a_i$ is associated with two payoff/cost processes $X^i$, $Y^i$. %and a terminal payoff/cost random variable $\xi^i$.
Their expected utilities $J_i(\t_1,\t_2)$, $i=1,2$, are defined as follows:
\be \label{J12}
\ba{lll} \dis J_1(\t_1,\t_2) \dfnn E\Big\{X^1_{\t_1}1_{\{\t_1\le \t_2\}} + Y^1_{\t_2}1_{\{\t_2< \t_1\}}\Big\}\\\mbox{ and }\\
\dis J_2(\t_1,\t_2) \dfnn
E\Big\{X^2_{\t_2}1_{\{\t_2<
\t_1\}}+ Y^2_{\t_1}1_{\{\t_1\le \t_2\}}\Big\}.
\ea \ee
That is, if the player $a_i$ is the one who actually stops the game (i.e. $\t_i<\t_j$ for $j\neq i$),
then he receives $X^i_{\t_i}$; if the game is stopped by the other player $a_j$ (i.e. $\t_j<\t_i$),
 then $a_i$ receives $Y^i_{\t_j}$.
%If nobody stops the game early (i.e. $\t_1=\t_2=T$), then the game expires anyway and $a_i$ receives $\xi_i$.
In the case that $\t_1=\t_2$ we take the convention that $a_1$ is
responsible for stopping the game. We can of course assume instead
that $a_2$ is responsible in this case and thus the corresponding
payoffs/costs inside the expectations in (\ref{J12}) become
$$
X^1_{\t_1}1_{\{\t_1< \t_2\}} + Y^1_{\t_2}1_{\{\t_2\le \t_1\}} ~~\mbox{and}~~ X^2_{\t_2}1_{\{\t_2\le
\t_1\}}+ Y^2_{\t_1}1_{\{\t_1< \t_2\}}.
$$
 %We can also consider some sharing rules in this case by considering, for example,
%${1\over 2}[X^1_{\t_1}+Y^1_{\t_1}]1_{\{\t_1=\t_2\}}$ and ${1\over 2}[X^2_{\t_1}+Y^2_{\t_1}]1_{\{\t_1=\t_2\}}$.

Throughout the paper we shall use the following assumptions.

\no{\bf A1.}  The processes $X^1, X^2, Y^1, Y^2$ belong to the space
[D], and $X^1, X^2$ have only positive jumps;

\no{\bf A2.}  P-a.s.,   $X^i_t \le Y^i_t$ for any $t\leq T$;

\no{\bf A3.}  For any $\t\in\cT_0$, $P(\{X^1_\t < Y^1_\t\}\backslash\{X^2_\t<Y^2_\t\}) =0$. % where $\D$ denotes the symmetric difference.

\ms The assumption {\bf A1} is more or less the minimum requirement
for the problem. {\bf A2} implies that there is penalty for stopping
the game early.
 We can study similarly the situation with reward for early stopping, namely to replace {\bf A2} with $X^i_t\ge Y^i_t$.
 Moreover, if we assume $X^2<Y^2$, then {\bf A3} is redundant.
\ms

%For any $\bF$-stopping times $\t_1, \t_2 \in \cT_0$,

%The game has two players $a_1$ and $a_2$. For $i=1,2$, $\t_i$ is the The quantities
%$J_1(\t_1,\t_2)$ and $J_2(\t_1,\t_2)$ stand for respectively the rewards
%of two players $a_1$ and $a_2$, who control a system, when they make
%the decision to stop controlling at $\t_1$ and $\t_2$ for $a_1$ and
%$a_2$ respectively.

Our main goal is to study the NEP of the game.
\begin{defn}
\label{equidefn} We say that $(\t_1^*,\t_2^*)\in {\cT_0}^2$ is a
Nash Equilibrium Point of the Nonzero-sum Dynkin game associated
with $J_1$ and $J_2$ if:
 \be
\label{equilibrium} J_1(\t_1,\t_2^*)\le J_1(\t_1^*,\t_2^*),\q
J_2(\t_1^*,\t_2)\le J_2(\t_1^*,\t_2^*),\q \forall \t_1,\t_2\in \cT_0. \ee
\end{defn}

As pointed out previously, this problem has been studied by
several authors in the Markovian framework \cite{bfried, catlep,
nagai, oht}, $i.e$, when besides to Assumptions {\bf A1}-{\bf A3}, the
processes $X^i$ and $Y^i$ are deterministic functions of a Markov
process $(m_t)_{t\leq T}$. If this latter condition is not
satisfied, E.Etourneau showed in \cite{etourneau} that the game has
a NEP when $Y^1$ and $ Y^2$ are supermartingales. Note that even in the Markovian
framework authors assume an equivalent condition to Etourneau's one.

Our main result is the following theorem, which assumes only
Assumptions {\bf A1}-{\bf A3} but without any regularity assumption
on $Y^1$, $Y^2$.

\begin{thm}
\label{existence} Under Assumptions {\bf A1}, {\bf A2} and {\bf A3},
the nonzero-sum Dynkin game associated with $J_1$ and $J_2$ has an
NEP $(\t_1^*,\t_2^*)$.
\end{thm}

We shall construct $(\t^*_1, \t^*_2)$ in next section.
%
%So the main
%objective of this article is to overcome this regularity assumption
%on $Y^1$, $Y^2$ and to prove that the nonzero-sum Dynkin game
%associated with $J_1$ and $J_2$ has a Nash equilibrium point under
%Assumptions \ref{assumption} only.
%
Our construction is based on the Snell envelope of processes which we
introduce briefly now. For more details on this subject one
can refer e.g. to El-Karoui \cite{Elka} or Dellacherie and Meyer
\cite{DM}. \bs

\begin{lem} (\cite{DM}, pp.431 or \cite{Elka}, pp.140) \label{thmsnell} Let $U$ be a process in the space [D].
%an
%$\bF$-adapted $\R$-valued RCLL process that belongs to the class
%[D], $i.e.$\@ the set of random variables $\{U_\tau, \,\,\tau
%\in{\cal T}_0\}$ is uniformly integrable.
Then, there exists an
$\bF$-adapted $\R$-valued RCLL process $W$
such that $W$ is the smallest super-martingale which dominates $U$,
$i.e$, if $\bar{W}$ is another RCLL
supermartingale such that  $\bar{W}_t\geq
U_t$ for all $0\leq t\leq T$, then $\bar{W}_t\geq W_t$ for any $0\leq t\leq T$. The process
$W$ is called the {\it Snell envelope\,} of $U$. Moreover, the
following properties hold:

$(i)$ For any $\bF$-stopping time $\theta$ we have:  \be \label{sun}
W_\theta=\esssup_{\tau \in {\cal
T}_{\theta}}E[U_\tau|\cF_\theta]\,\,\,\,\,\,(\mbox{and then
}W_T=U_T), ~~ P-a.s.\ee

$(ii)$ Assume that $U$ has only positive jumps. Then the stopping
time  $$\tau^*\dfnn\inf\{s\geq 0, W_s=U_s\}\wedge T$$ is optimal,
$i.e.$,
\begin{equation}\label{sdeux}E[W_0]=E[W_{\tau^*}]=E[U_{\tau^*}]=\sup_{\tau \in\cT_0}E[U_\tau].\end{equation}
\end{lem}

\begin{rem}\label{snelmart} As a by-product of (\ref{sdeux}) we have
$W_{\tau^*}=U_{\tau^*}$ and the process $W$ is a martingale on the
time interval $[0,\tau^*]$.
\end{rem}

\section{Construction of a Nash Equilibrium Point}
\setcounter{equation}{0}

In this section we shall construct a sequence of pairs of decreasing
stopping times $(\t_{2n+1}, \t_{2n+2})$ and show that their limits
$(\t^*_1, \t^*_2)$ is an NEP.  First, notice that $Y^1$ is only
required to be RCLL, and that $Y^1_T$ is never used in (\ref{J12}),
for notational simplicity at below we will also assume without loss
of generality that

\ms

\no{\bf A4.} P-a.s., $Y^1_T = X^1_T$.

\ms

\no We emphasize again that this is just for notational simplicity.
Without assuming {\bf A4}, we may replace the integrands in
(\ref{Wn1}) below with
 $$
 X^1_\t 1_{\{\t< \t_{2n}\}} +
\Big[X^1_T 1_{\{\t_{2n}=T\}} +
Y^1_{\t_{2n}}1_{\{\t_{2n}<T\}}\Big]1_{\{\t\ge\t_{2n}\}},
$$
 and all the arguments will be the same.
%Throughout the section Assumption \ref{assumption} is always in force.

%\noindent {\it Proof}: it will be given in four steps. \bs

%\noindent \underline{{\bf Step 1}}: construction of the
%approximating sequence of stopping times and the Nash point. \bs

\bs
 We start with defining $\t_1\dfnn T$ and $\t_2\dfnn T$.  For
$n=1, \cds$, assume $\t_{2n-1}$ and $\t_{2n}$ have been defined, we
then define $\t_{2n+1}$ and $\t_{2n+2}$ as follows. First, let
 \be \label{Wn1}
 W^{2n+1}_t \dfnn \esssup_{\t\in\cT_t}E_t\Big\{X^1_\t 1_{\{\t< \t_{2n}\}}
+ Y^1_{\t_{2n}}1_{\{\t\ge\t_{2n}\}}\Big\},\q t\leq T;
  \ee
where and in the sequel $E_t\{\cd\}\dfnn E\{\cd|\cF_t\}$,   and
 \be \label{tn1}
 \tilde \t_{2n+1} \dfnn \inf\{t\ge 0: W^{2n+1}_t = X^1_t\}\wedge
 \t_{2n};\q \t_{2n+1}\dfnn
 \left\{\ba{lll}
 \tilde \t_{2n+1},\q{\rm if}~~ \tilde \t_{2n+1}<\t_{2n};\\
 \t_{2n-1},\q{\rm if}~~ \tilde \t_{2n+1}=\t_{2n}. \ea\right.
 \ee
 Next, let
 \be \label{Wn2}
 W^{2n+2}_t \dfnn \esssup_{\t\in\cT_t}E_t\Big\{X^2_\t 1_{\{\t< \t_{2n+1}\}}
+ Y^2_{\t_{2n+1}}1_{\{\t\ge\t_{2n+1}\}}\Big\},\q t\leq T;
  \ee
  and
 \be \label{tn2}
 \tilde \t_{2n+2} \dfnn \inf\{t\ge 0: W^{2n+2}_t = X^2_t\}\wedge
 \t_{2n+1};\q \t_{2n+2}\dfnn
 \left\{\ba{lll}
 \tilde \t_{2n+2},\q{\rm if}~~ \tilde \t_{2n+2}<\t_{2n+1};\\
 \t_{2n},\q{\rm if}~~ \tilde \t_{2n+2}=\t_{2n+1}. \ea\right.
 \ee
  We note that the integrand in (\ref{Wn1}) is slightly
 different from that of $J_1(\t, \t_{2n})$ in (\ref{J12}). The main reason is that, in order to apply Lemma \ref{thmsnell},
 we need the process $U^{2n+1}$ in (\ref{U1n}) below to be RCLL. But nevertheless we will prove
 later in Lemma \ref{localopt} that $W^{2n+1}$ serves our purpose well.

\begin{lem}
\label{tn}
Assume Assumptions {\bf A1} and {\bf A2}. For $n=1, 2, \cds$, $\t_{n}$ is a stopping time and $\t_{n+2}\le \t_{n}$.
\end{lem}
{\it Proof.} We prove the following stronger results  by induction
on $n$:
 \be \label{induction}
 \t_{n}\in \cT_0,\q \{\t_{n}<\t_{n+1}\}\subset \{\tilde\t_{n+2} \le \t_{n}\},\q \t_{n+2}\le \t_{n}.
% \label{induction2}
% && \t^{n-1}_2 \mbox{is a stopping times},
%~~ \{\t^{n-1}_2<\t^{n-1}_1\}\subset \{\tilde\t^n_2 \le \t^{n-1}_2\},~~  \t^n_2\le \t^{n-1}_2.
%&& \{\t^{n-1}_1<\t^{n-1}_2\}\subset \{\tilde\t^n_1 \le \t^{n-1}_1\},~~ \{\t^{n-1}_2<\t^n_1\}\subset \{\tilde \t^n_2\le \t^{n-1}_2\};\\
% \label{induction3}
%&& \{\t^{n-1}_2=\t^{n-1}_1\}\subset \{\t^{n-1}_1=T\}, ~~ \{\t^n_1 =
%\t^{n-1}_2\} \subset \{\t^{n-1}_2=T\}.
 \ee

Obviously (\ref{induction}) holds for $n=1,2$. Assume it is true for $2n-1$ and $2n$. We shall prove it for $2n+1$ and $2n+2$.

First, define
 \be
 \label{U1n}
 U^{2n+1}_t\dfnn X^1_t 1_{\{t< \t_{2n}\}} + Y^1_{\t_{2n}}1_{\{t\ge \t_{2n}\}}.
\ee Since $\t_{2n}$ is a stopping time, by Assumptions {\bf A1} and
{\bf A2} we know $U^{2n+1}$ is in space [D] and has only positive
jumps. Apply Lemma \ref{thmsnell},
% and recall in particular Assumption {\bf A2} (in order to apply Lemma \ref{thmsnell} (ii)) ,
$W^{2n+1}$ is the snell envelope of $U^{2n+1}$ and $\tilde
\t_{2n+1}$ is the optimal stopping time.

If $\t_{2n-1} < \t_{2n}$, then by the second claim of (\ref{induction}) for $2n-1$ we
have $\tilde \t_{2n+1} \le \t_{2n-1}$ and thus $\tilde \t_{2n+1} <
\t_{2n}$. This implies that
 \be
 \label{ind3}
 \{\tilde \t_{2n+1} = \t_{2n}\}\subset \{\t_{2n-1}\ge \t_{2n}\},
 \ee
  which, combined with the
definition (\ref{tn1}), implies further that $\t_{2n+1}$ is a stopping
time.

Next, on $\{\t_{2n+1}<\t_{2n+2}\}$, by definition of $\t_{2n+2}$ in
(\ref{tn2}) we have $\t_{2n+2}=\t_{2n}$. Then $U^{2n+3}_t = U^{2n+1}_t$ for $t\ge \t_{2n+1}$ and thus
 \be
 \label{ind4}
W^{2n+3}_{\t_{2n+1}} 1_{\{\t_{2n+1}<\t_{2n+2}\}} = W^{2n+1}_{\t_{2n+1}}
1_{\{\t_{2n+1}<\t_{2n+2}\}}.
 \ee
On the other hand, if $\tilde \t_{2n+1} = \t_{2n}$, by
the third claim of (\ref{induction}) for $2n$, (\ref{ind3}), and definition of
(\ref{tn1}), we have $\t_{2n+2} \le \t_{2n} \le \t_{2n-1} =
\t_{2n+1}$. Thus $\{\t_{2n+1} <\t_{2n+2}\} \subset \{\t_{2n+1} = \tilde \t_{2n+1} <
\t_{2n}\}$, and therefore, by Remark \ref{snelmart},
$$
 W^{2n+1}_{\t_{2n+1}} 1_{\{\t_{2n+1}<\t_{2n+2}\}} = X^1_{\t_{2n+1}}
1_{\{\t_{2n+1}<\t_{2n+2}\}}.
$$
 This, together with (\ref{ind4}), implies that
 $$
  W^{2n+3}_{\t_{2n+1}} 1_{\{\t_{2n+1}<\t_{2n+2}\}} = X^1_{\t_{2n+1}}
1_{\{\t_{2n+1}<\t_{2n+2}\}}.
$$
Now by the definition of $\tilde\t_{2n+3}$ in (\ref{tn1}) we know
 \be \label{ind5}
  \{\t_{2n+1}<\t_{2n+2}\}\subset \{\tilde\t_{2n+3} \le
\t_{2n+1}\}.
 \ee

 Moreover, if $\t_{2n+3} >\t_{2n+1}$, by definition (\ref{tn1})
we have $\t_{2n+3}=\tilde \t_{2n+3} < \t_{2n+2}$. Then
$\t_{2n+1}<\tilde\t_{2n+3}<\t_{2n+2}$. This contradicts with
(\ref{ind5}). Therefore, $\t_{2n+3}\le \t_{2n+1}$.

Finally, one can prove (\ref{induction}) for $2n+2$ similarly. \qed

Following is another important property of the stopping times
$\t_n$.
\begin{lem}
\label{tnT} Assume Assumptions {\bf A1} and {\bf A2}. On $\{\t_n = \t_{n-1}\}$, we have $\t_m  =T$ for all $m\le n$.
 \end{lem}
  {\it Proof.} The result is obvious for $n=2$. Assume it is true for $n$.
  Now for $n+1$, on  $\{\t_{n+1} = \t_{n}\}$,
  by the definition of $\t_{n+1}$ in (\ref{tn1}) or (\ref{tn2}) we have
  $\t_{n+1}=\t_{n-1}$. Then $\t_n=\t_{n-1}$ and thus by induction
  assumption we get the result.
\qed

Next lemma shows that $\t_n$ is the optimal stopping time for some problem.

\begin{lem}
\label{localopt}
 Assume Assumptions {\bf A1}, {\bf A2} and {\bf A4}. For any $\t\in \cT_0$ and any $n$ we have:
 \be \label {opt}
 J_1(\t,\t_{2n}) \le J_1(\t_{2n+1}, \t_{2n})~~ \mbox{ and } ~~J_2(\t_{2n+1},\t)\le J_2(\t_{2n+1}, \t_{2n+2}).
 \ee
 \end{lem}
 {\it Proof.} First, by the definition of $W^{2n+1}$ in (\ref{Wn1}) we have $W^{2n+1}_{\t_{2n}} = Y^1_{\t_{2n}}$.  Next, by Lemma \ref{thmsnell} we have  $W^{2n+1}_t \ge X^1_t$ for any $t\in [0,\t_{2n}]$ and $W^{2n+1}$ is
a supermartingale over $[0,\t_{2n}]$. Then, for any $\t\in\cT_0$,
 \bea
 \label{T2n1opt}
 && J_1(\t,\t_{2n})= E\Big\{X^1_{\t}1_{\{\t\le \t_{2n}\}} +
Y^1_{\t_{2n}} 1_{\{\t_{2n}<\t\}}\Big\}\\
 &&\le E\Big\{W^{2n+1}_{\t}1_{\{\t\le \t_{2n}\}} + W^{2n+1}_{\t_{2n}} 1_{\{\t_{2n}<\t\}}\Big\}=E\{W^{2n+1}_{\t_{2n}\wedge\t}\}\le W^{2n+1}_0.\nonumber
 \eea
On the other hand, by Lemma \ref{tnT} and Assumption {\bf A4} we have
 \beaa
J_1(\t_{2n+1}, \t_{2n})&=&E\Big\{X^1_{\t_{2n+1}}1_{\{\t_{2n+1}\le \t_{2n}\}} + Y^1_{\t_{2n}} 1_{\{\t_{2n}<\t_{2n+1}\}}\Big\}\\
&=&E\Big\{X^1_{\t_{2n+1}}1_{\{\t_{2n+1}< \t_{2n}\}} + Y^1_{\t_{2n}} 1_{\{\t_{2n}\le \t_{2n+1}\}}\Big\}.
 \eeaa
 By (\ref{tn1}), (\ref{ind3}), and then by Remark \ref{snelmart}, we get
 $$
  J_1(\t_{2n+1}, \t_{2n}) = E\Big\{X^1_{\tilde\t_{2n+1}}1_{\{\tilde \t_{2n+1}< \t_{2n}\}} + W^{2n+1}_{\t_{2n}} 1_{\{\tilde\t_{2n+1}= \t_{2n}\}}\Big\}=E\{W^{2n+1}_{\tilde \t_{2n+1}}\}= W^{2n+1}_0.
  $$
 This, together with (\ref{T2n1opt}), proves $J_1(\t,\t_{2n}) \le J_1(\t_{2n+1}, \t_{2n})$.

 Similarly we can prove $J_2(\t_{2n+1},\t)\le J_2(\t_{2n+1}, \t_{2n+2})$.
 \qed

 \bs

 Now define
 \be \label{T12*}
 \t_1^* \dfnn \lim_{n\to\infty} \t_{2n+1} ~~\mbox{ and } ~~ \t_2^* \dfnn\lim_{n\to\infty} \t_{2n}.
 \ee
 We shall prove that $(\t^*_1, \t^*_2)$ is an NEP.  We divide the proof into several lemmas.

 \begin{lem}
 \label{Jtn}
 Assume Assumptions {\bf A1} and {\bf A2}.

 (i) For any $\t\in \cT_0$, we have $\dis\lim_{n\to\infty} J_1(\t, \t_{2n}) = J_1(\t, \t^*_2)$.

 (ii) For any $\t\in \cT_0$ such that $P(\t=\t^*_1<T)=0$, we have $\dis\lim_{n\to\infty} J_2(\t_{2n+1}, \t) = J_2(\t^*_1, \t)$.
 \end{lem}
 {\it Proof.} (i) By Assumption {\bf A1}, we have
 \beaa
\lim_{n\to\infty}J_1(\t,\t_{2n}) &=& \lim_{n\to\infty}E\Big\{X^1_{\t}1_{\{\t\le \t_{2n}\}} + Y^1_{\t_{2n}} 1_{\{\t_{2n}<\t\}}\Big\}\\
 &=& E\Big\{X^1_{\t}1_{\{\t\le \t_2^*\}} + Y^1_{\t_2^*} 1_{\{\t_2^*<\t\}}\Big\}
=J_1(\t, \t_2^*).
\eeaa

(ii) Since $\{\t<\t_{2n+1}\}\subset \{\t<T\}$, we have
$$
\lim_{n\to\infty}E\Big\{X^2_{\t} 1_{\{\t<\t_{2n+1}\}}\Big\}=\lim_{n\to\infty}E\Big\{X^2_{\t} 1_{\{\t<\t_{2n+1}, \t \neq \t^*_1\}}\Big\} =E\Big\{X^2_{\t} 1_{\{\t<\t_1^*\}}\Big\}.
$$
Moreover, note that $\t^*_1\le \t_{2n+1}$, then $\{\t^*_1=T\}\subset\{\t_{2n+1}=T\}$. Applying the assumption $P(\t=\t^*_1<T)=0$ twice we have
\beaa
&&\lim_{n\to\infty}E\Big\{Y^2_{\t_{2n+1}}1_{\{\t_{2n+1}\le \t\}}\Big\}=\lim_{n\to\infty}E\Big\{Y^2_{\t^*_1}1_{\{\t_{2n+1}\le \t\}}\Big[1_{\{\t\neq\t^*_1\}} + 1_{\{\t=\t^*_1\}}\Big]\Big\}\\
&&=E\Big\{Y^2_{\t^*_1}1_{\{\t^*_1< \t\}} + Y^2_{\t^*_1}1_{\{\t=\t^*_1=T\}}\Big\} = E\Big\{Y^2_{\t^*_1}1_{\{\t^*_1\le \t\}} \Big\}.
\eeaa
Then
\beaa
\lim_{n\to\infty}J_2(\t_{2n+1}, \t)&=&\lim_{n\to\infty}E\Big\{X^2_{\t} 1_{\{\t<\t_{2n+1}\}}+ Y^2_{\t_{2n+1}}1_{\{\t_{2n+1}\le \t\}}\Big\}\\
&=& E\Big\{X^2_{\t} 1_{\{\t<\t_1^*\}} + Y^2_{\t_1^*}1_{\{\t_1^*\le \t\}}\Big\} = J_2(\t_1^*,\t).
\eeaa
The proof is complete.
\qed

 \begin{lem}
 \label{Jt*}
 Assume Assumptions {\bf A1}-{\bf A4}. Then it holds that
$$
\lim_{n\to\infty} J_1(\t_{2n+1}, \t_{2n}) = J_1(\t^*_1, \t^*_2);\q
\lim_{n\to\infty} J_2(\t_{2n-1}, \t_{2n}) = J_2(\t^*_1, \t^*_2).
$$
%\bea
%\label{PT2n1}
%&&\lim_{n\to\infty} P\Big(Y^1_{\t^*_1}>X^1_{\t^*_1},
%\t_{2n+1}\le \t_{2n}, \t^*_1=\t^*_2\Big) = 0;\\
%\label{PT2n2}
%&&\lim_{n\to\infty} P\Big(Y^2_{\t^*_1}>X^2_{\t^*_1},
%\t_{2n}< \t_{2n-1}, \t^*_1=\t^*_2\Big) = 0.
%\eea
\end{lem}
{\it Proof.} (i) We first show that
\be
\label{Jt*2}
\lim_{n\to\infty} J_2(\t_{2n-1}, \t_{2n}) = J_2(\t^*_1, \t^*_2).
\ee
Note that
$$
J_2(\t_{2n-1}, \t_{2n})=E\Big\{\Big[X^2_{\t_{2n}}1_{\{\t_{2n}< \t_{2n-1}\}} + Y^2_{\t_{2n-1}}
1_{\{\t_{2n-1}\le \t_{2n}\}}\Big]\Big[1_{\{\t_1^*\neq\t^*_2\}} + 1_{\{\t^*_1=\t^*_2\}}\Big]\Big\}.
$$
Since $X^2,Y^2$ are in space [D], sending $n\to\infty$ we have
\bea
\label{J2n1}
&&\lim_{n\to\infty} J_2(\t_{2n-1}, \t_{2n})\nonumber\\
&&=\lim_{n\to\infty}E\Big\{X^2_{\t^*_2}1_{\{\t^*_2<\t^*_1\}} + Y^2_{\t^*_{1}} 1_{\{\t^*_1< \t^*_2\}}+\Big[X^2_{\t^*_2}1_{\{\t_{2n}< \t_{2n-1}\}}+Y^2_{\t^*_1}
1_{\{\t_{2n-1}\le \t_{2n}\}}\Big]1_{\{\t^*_1=\t^*_2\}} \Big\}\nonumber\\
 &&=E\Big\{X^2_{\t^*_2}1_{\{\t^*_2<\t^*_1\}} + Y^2_{\t^*_{1}} 1_{\{\t^*_1\le \t^*_2\}}\Big\} + I = J_2(\t^*_1, \t^*_2) + I,
 \eea
 where
 \be
 \label{I}
 I\dfnn \lim_{n\to\infty}E\Big\{\Big[X^2_{\t^*_1}-Y^2_{\t^*_1}\Big] 1_{\{\t_{2n}< \t_{2n-1},\t^*_1=\t^*_2\}}\Big\}.
\ee

On the other hand, set
$$
\t\dfnn\left\{\ba{lll}
\t_2^*,\q {\rm if}~~ \t_2^*<\t_1^*;\\
T,\q{\rm if}~~ \t_2^*\ge \t_1^*. \ea\right.
$$
Then $\t\in\cT_0$ and $P(\t = \t^*_1<T)=0$. By Lemma
\ref{Jtn} (ii) we have \beaa
&&\lim_{n\to\infty} J_2(\t_{2n-1}, \t) = J_2(\t^*_1,\t) =E\Big\{ X^2_{\t} 1_{\{\t< \t^*_1\}}
+ Y^2_{\t^*_1}1_{\{{\t}\ge \t^*_1\}}\Big\}\\
&&=E\Big\{X^2_{\t^*_2}1_{\{\t^*_2<\t^*_1\}} + Y^2_{\t^*_{1}}
1_{\{\t^*_1\le \t^*_2\}}\Big\} = J_2(\t^*_1, \t^*_2). \eeaa By Lemma
\ref{localopt}, we get $I\ge 0$. Now by Assumption {\bf A2} we have
\be \label{I=0} I=0. \ee Then (\ref{J2n1}) implies (\ref{Jt*2}).

(ii) It remains to prove
\be
\label{Jt*1}
\lim_{n\to\infty} J_1(\t_{2n+1}, \t_{2n}) = J_1(\t^*_1, \t^*_2).
\ee
Similar to (\ref{J2n1}) we have
\bea
&&\lim_{n\to\infty} J_1(\t_{2n+1}, \t_{2n})\nonumber\\
&& = \lim_{n\to\infty}E\Big\{X^1_{\t^*_1}1_{\{\t^*_1<\t^*_2\}} + Y^1_{\t^*_{2}}
1_{\{\t^*_2<\t^*_1\}} + \Big[X^1_{\t^*_1} 1_{\{\t_{2n+1}\le \t_{2n}\}} + Y^1_{\t^*_1} 1_{\{\t_{2n+1}> \t_{2n}\}}\Big]1_{\{\t^*_1=\t^*_2\}}\Big\}\nonumber\\
&&= \lim_{n\to\infty}E\Big\{X^1_{\t^*_1}1_{\{\t^*_1\le \t^*_2\}} +
Y^1_{\t^*_{2}} 1_{\{\t^*_2<
\t^*_1\}}+\Big[Y^1_{\t^*_1}-X^1_{\t^*_1}\Big] 1_{\{\t_{2n+1}>
\t_{2n},\t^*_1=\t^*_2\}}\Big\}\nonumber\\
\label{J1n1} &&= J_1(\t^*_1, \t^*_2) +
\lim_{n\to\infty}E\Big\{\Big[Y^1_{\t^*_1}-X^1_{\t^*_1}\Big]
1_{\{\t_{2n+1}> \t_{2n},\t^*_1=\t^*_2\}}\Big\}. \eea

By Assumption {\bf A2}, we get from (\ref{I=0}) that
 $$
 \lim_{n\to\infty} P\Big(X^2_{\t^*_1}<Y^2_{\t^*_1}, \t_{2n+2}<
 \t_{2n+1},\t^*_1=\t^*_2\Big) = 0.
 $$
Applying the third claim in Lemma \ref{tn} we have
$\{\t_{2n}<\t_{2n+1}\}\subset \{\t_{2n+2}<\t_{2n+1}\}$. Then by
Assumption {\bf A3} we have
 $$
 \lim_{n\to\infty} P\Big(X^1_{\t^*_1}<Y^1_{\t^*_1}, \t_{2n}<
 \t_{2n+1},\t^*_1=\t^*_2\Big) = 0.
 $$
Then (\ref{J1n1}) leads to (\ref{Jt*1}) immediately. \qed

We are now ready to show that $(\t^*_1, \t^*_2)$ is an NEP. \ms

{\it Proof of Theorem \ref{existence}.} We recall again that
Assumption {\bf A4} is just for notational simplicity. So in the
proof we may assume it.

First, by Lemma \ref{Jtn} (i), Lemma \ref{Jt*}, and Lemma
\ref{localopt}, we have \be \label{J1*} J_1(\t, \t^*_2) \le
J_1(\t^*_1, \t^*_2), \q\forall \t\in \cT_0. \ee

 Similarly, for any $\t$ such that $P(\t= \t_1^*<T)=0$, we have
\be
\label{J2*}
 J_2(\t_1^*,\t)\le J_2(\t_1^*,\t_2^*).
 \ee
In the general case, for any $\t\in\cT_0$,  set
$$
\hat \t_n\dfnn \left\{\ba{lll}
[\t+{1\over n}]\wedge T,\q {\rm if}~~ \t= \t_1^*<T;\\
\t,\q {\rm otherwsie}. \ea\right.
$$
Then $\hat\t_n$ is a stopping time and $P(\hat\t_n= \t_1^*<T)=0$.  Thus (\ref{J2*}) leads to
$$
J_2(\t_1^*,\hat\t_n)\le  J_2(\t_1^*,\t_2^*).
$$
Send $n\to\infty$, we obtain (\ref{J2*}) for general $\t$.

Combine (\ref{J1*}) and (\ref{J2*}),  we obtain $(\t_1^*,\t_2^*)$ is
an NEP. \qed
\begin{remark} In the case when $X_2=-Y_1$ and $Y_2=-X_1$ then $J_1+J_2=0$, i.e. we fall in the
framework of the well known zero-sum Dynkin game and then the NEP
for the game is just a saddle-point. Comparing to the result by
Lepeltier and Mainguenau \cite{lepmaing}, which is the most general
paper on this subject known to date, our result provides a new
construction method of the saddle point. Additionally it is obtained
under less regularity conditions on the processes $X_1$ and $X_2$.
\qed
\end{remark}

\section{Application to game contingent claims}
\setcounter{equation}{0}

It is by now well-known that an American contingent claim is a
contract which allows its holder to exercise at a time she decides
before or at the maturity. The only role of its issuer is to
provide, if any, the pledged wealth to the buyer. In contrary, an
American game contingent claim (ACC for short) is mainly an American
contingent claim where the issuer is also allowed to recall/cancel
the contract. Actually assume that $a_1$ (resp. $a_2$) is the issuer
(resp. buyer) of the ACC. Both sides are allowed to exercise.
Therefore it enables $a_1$ to terminate it and $a_2$ to exercise it
at any time up to maturity date $T$ when the contract is expired
anyway. Also if $a_2$ decides to exercise at $\sigma$ or $a_1$ to
terminate at $\tau$ then $a_1$ pays to $a_2$ the amount:
$$\Gamma(\tau,\sigma)=L_\sigma 1_{[\sigma\leq \tau, \si<T]}+U_\tau
1_{[\tau< \sigma]}+\xi 1_{[\tau= \sigma=T]}$$ where:

- $\sigma$ and $\tau$ are two $\bF$-stopping times

- $L$ and $U$ are $\bF$-adapted continuous processes such that
$L\leq U$. The quantity $L_\sigma$ (resp. $U_\tau$) is the amount
that obtains $a_2$ (resp. pays $a_1$) for her decision to exercise
(resp. cancel) first at $\sigma$ (resp. $\tau$). The difference
$U-L$ represents the compensation that $a_1$ pays to $a_2$ for the
decision to terminate the contract before maturity date $T$

- $\xi$ is an $\cF_T$-random variable which satisfies $L_T\leq \xi
\leq U_T$. It stands for the money that $a_1$ pays to $a_2$ if both
accept to terminate the GCC at maturity date $T$.

For this contingent claim, the seller $a_1$ (resp. buyer $a_2$) aims
at maximizing (resp. minimizing) her cost (resp. reward) in
expectation, $i.e.$, the quantity:
$$J(\tau,\sigma):=E[\Gamma(\tau,\sigma)].$$
where $E[.]$ is the expectation under the probability $P$ on the
space $(\Omega, \cF)$. \ms

Game contingent claims are introduced by Y.Kifer in \cite{kifer} in
the framework of the Black and Scholes model. Since then, there have
been several papers on the same subject \cite{bk, hamadene, kkal}.
In a complete market, it is shown in those works that the
non-arbitrage price $V_0$ of the GCC is equal to the zero-sum Dynkin
game associated with $L$ and $U$, i.e., $$ V_0=\esssup_{\sigma \geq
0}\essinf_{\tau\geq 0}J(\t,\sigma)=\essinf_{\tau\geq
0}\esssup_{\sigma \geq 0}J(\t,\sigma).$$

Another point of view for pricing American game options, especially
in incomplete markets and in connection with the utility
maximization approach, is introduced by C.Kuhn in \cite{kuhn2} and
which is the following: \ms

Let $\varphi_1,\varphi_2: \R\rightarrow \R$ be non-decreasing and
concave functions. Those functions stand for utility functions of
the seller, respectively, the buyer of the GCC. The seller $a_1$
(resp. the buyer $a_2$) chooses a stopping time $\tau$ (resp.
$\sigma$) in order to maximize
$$J_1(\tau,\sigma):=E[\varphi_1(-\Gamma(\tau,\sigma))] \,\,(\mbox{resp.
}J_2(\tau,\sigma):=E[\varphi_2(\Gamma(\tau,\sigma))]).$$ Therefore
if the nonzero-sum Dynkin game associated with $J_1$ and $J_2$ has a
Nash equilibrium point $(\sigma^*,\tau^*)$, i.e.,
$$J_1(\tau^*,\sigma^*)\geq J_1(\tau,\sigma^*) \mbox{ and }J_2(\tau^*,\sigma^*)\geq J_2(\tau^*,\sigma)$$ then
$-\f_1^{-1}(J_1(\tau^*,\sigma^*))$ (resp. $\f_2^{-1}(J_2(\tau^*,\sigma^*))$) is a seller
(resp. buyer) price of the GCC. \bs

Note that if $\varphi_1(x)=\varphi_2(x)=x, \forall x\in \R$, i.e.
the agents $a_1$ and $a_2$ are risk-neutral, then the nonzero-sum
game is actually a zero-sum Dynkin game, $(\tau^*,\sigma^*)$ is a
saddle-point for this game and
$-J_1(\tau^*,\sigma^*)=J_2(\tau^*,\sigma^*)$. Moreover this latter
quantity is the value of the game. For more details on zero-sum
Dynkin games one can see e.g. \cite{nazaret2, bfried2, CK, hamadene,
ls, lepmaing, stet, tv}. \ms

So pricing the GCC described above turns into the existence of a NEP
for the associated nonzero-sum Dynkin game. In \cite{kuhn2}, based
on the article by Morimoto \cite{morimoto2}, the author has just
been able to show the existence of that NEP in the set of discrete
stopping times and exponential utility functions. Also using the
result of the previous section, we are able to fill in the gap
between the discrete stopping times used in \cite{kuhn2} and
continuous ones which we use here and, on the other hand, to allow
for arbitrary utility functions for the agents. Actually we have:
\begin{thm}
Assume that:

(i) The utility functions $\varphi_1$ and $\varphi_2$ are
non-decreasing;

(ii) $L_t \le U_t$ and $L_T\le \xi \le U_T$, P-a.s.;

(iii) The processes $\f_1(-L), \f_1(-U), \f_2(L), \f_2(U)$ are in
the space [D]; and the random variables $\f_1(-\xi)$ and $\f_2(\xi)$
are square integrable.

(iv) The processes $\f_1(-U)$ and $\f_2(L)$ has only positive jumps.

 Then the nonzero-sum Dynkin game associated with the GCC has a
Nash equilibrium point $(\tau^*,\sigma^*)$.
\end{thm}
{\it Proof}: Define
 \beaa
 && X^1_t \dfnn \f_1(-U_t)1_{\{t<T\}} + \f_1(-\xi)1_{\{t=T\}},\q
 X^2_t \dfnn \f_2(L_t)1_{\{t<T\}} + \f_2(\xi)1_{\{t=T\}};\\
 && Y^1_t \dfnn \f_1(-L_t)1_{\{t<T\}} + \f_1(-\xi)1_{\{t=T\}},\q
 Y^2_t \dfnn \f_2(L_t)1_{\{t<T\}} + \f_2(\xi)1_{\{t=T\}}.
 \eeaa
 One can check straightforwardly that $X^1, Y^1, X^2, Y^2$
 satisfy Assumptions {\bf A1}-{\bf A4}, and that the value functions $J_1(\t, \si)$ and $J_2(\t, \si)$ are the same as those defined in (\ref{J12}).
 Then by Theorem \ref{existence}
   we obtain the desired result. $\qed$

\end{document}